\documentclass[11pt]{article}
\usepackage{amsmath}
\usepackage{fullpage}
\usepackage{textcomp}
\usepackage{graphicx}
\usepackage{sidecap}
\usepackage{setspace}
\usepackage{sectsty}
\usepackage{enumerate}
\usepackage{placeins}
\allsectionsfont{\normalfont\sffamily\bfseries}

\title{\sffamily\bfseries Towards a fully predictive model of flight paths in pigeons navigating in the familiar area: prediction across differing individuals}
\author{Richard P Mann \\
{\small School of Mathematics, University of Leeds, LS2 9JT}}
\date{}
\begin{document}

\maketitle

\section*{Introduction}
There is strong evidence that pigeons use visual cues to navigate within the familiar area \cite{braithwaite1991vfl}, and that visual landmarks constitute an important part of the familiar area map \cite{mann2011oil}. However, the precise role of the visual landscape in familiar area navigation remains controversial. In particular, the strongest evidence for visual landmark use -- the development and recapitulation of idiosyncratic habitual homing routes -- has been repeatedly observed in some locations \cite{biro2004frl,meade2005hpd,mann2011oil}, but not in others \cite{wiltschko2007hfp}. There is clear evidence that the specific nature of the release site influences whether pigeons will form familiar routes, and how idiosyncratic these may be \cite{schiffner2014pigeon}. The previous familiarity of birds with the landscape also effects whether the \emph{emergence} of habitual routes is clearly observable (see e.g. \cite{wiltschko2007hfp}). Some release sites are associated with a clear set of possible `corridors' for pigeons to learn, whereas others seem to lack this feature \cite{mann2011oil,schiffner2014pigeon}

Some specific visual cues have been implicated in the homing process. Roads, rivers and other strong linear features appear to be particularly attrative to pigeons even on their first flights from unfamiliar sites, and habitual routes are often formed over these \cite{biro2004frl,lipp2004pha}. Technical analysis of the landscape as presented in aerial photographs suggests that `edges' (areas where the image changes abruptly) are associated with changes in navigational behaviour \cite{lau2006aed}. Recent work \cite{mann2014landscape} has shown that pigeons' fidelity to their previous routes is connected to the edge density of the underlying landscape, suggesting a link between visual \emph{complexity} and navigational success, with optimal complexities found on the borders of forests and urban dwellings, as well as near roads and rivers. Visual inspection of candidate landmark sites suggest that other visually salient features such as church spires are likely to be used as visual cues as well \cite{mann2011oil}. 

All this evidence points to an emerging picture where certain types of landscape are more likely to be used as visual cues, leading to site-specific results when evaluating landscape use. The evidence so far suggests that strong linear features and the boundaries between different landscape types are particularly salient, and that pigeons have difficulty memorising landmarks in either visually barren (such as empty fields) or visually complex (such as inner urban or forest) landscapes. Nonetheless, because it is generally impossible to experimentally manipulate the landscape on a large scale, a complete understanding of how pigeons use visual cues has remained out of reach. It remains difficult to disentangle causality from correlation in these findings. However, while the landscape can not be manipulated directly, it is possible to choose new release sites to test specific hypotheses about the use of visual cues, \emph{if} those hypotheses make sufficiently precise predictions about navigational behaviour. For this reason we propose to move towards a predictive model of flight paths based on the visual landscape, where specific hypotheses about the mechanisms of visual cue use lead to quantitative predictions of expected flight routes, and can thus be tested against each other using data from recorded flights via statistical model selection (e.g. \cite{mann2013msi,mann2013tdo,mann2013humbugs}) 

This paper will detail the basis of our previously developed predictive model for pigeon flight paths based on observations of the specific individual being predicted. We will then describe how this model can be adapted to predict the flight of a new, unobserved bird, based on observations of other individuals from the same release site. We will test the accuracy of these predictions relative to naive models (no previous flight information) and those trained on the focal bird's own previous flights, and discuss the implications of these results for the nature of navigational cue use in the familiar area. Finally we will discuss how visual cues may be explicitly encoded in the model in future work.


\section*{A generative model of flight paths}
With GPS tracking technology, flight paths are now quantitative items of data, represented by the geographical position of the navigating bird over a series of time points. Many studies have focused on extracting patterns and/or statistics from these paths that can be used as the basis for further analysis, with examples including the tortuosity of the path, the fractal dimension, the distance between successive flights etc. However, in \cite{mann2011oil} we argued that we should model the flight path \emph{itself}, rather than extracted statistics. To do this we need to be able to discuss the \emph{probability} of an observation, i.e. of a specific flight path. 

We developed a generative model for flight paths \cite{mann2011oil,reece2011gps} based on the framework of Gaussian processes (GPs) \cite{rasmussen2006gpf}. GPs are a class of highly flexible probability distributions over functions, such as the position of a bird as a function of time. The reader is referred to \cite{rasmussen2006gpf} for a full introduction to the technical details of Gaussian process modeling, and to \cite{mann2011oil} for previous applications in pigeon navigation.

Taking the x-coordinate of the bird's position over time for example, we specified this as a random sample from a GP:
\begin{equation}
x(t) \sim \mathcal{GP}(h(t), k_\theta(t, t')),
\end{equation}
where $h(t)$ represents the \emph{habitual route} of the individual, and $k_\theta(t, t')$ is a covariance kernel function that controls the variability of flights. The habitual route itself is drawn from a different GP:
\begin{equation}
h(t) \sim \mathcal{GP}(s(t), k_\phi(t, t')),
\end{equation}
where $s(t)$ is the straight line path between the release point and home, and $k_phi(t, t')$ is another kernel controlling how variable habitual routes are between birds. 

A pair of paths can either be independent draws from the same GP, or can be dependently linked through a shared (unobserved) habitual route. In the case of the latter they collectively form a random sample from a higher dimensional GP:
\begin{equation}
X = [x_1(t_1); x_2(t_2)] \sim \mathcal{GP}([s(t_1); s(t_2)], \Sigma)
\end{equation}
where,
\begin{equation}
\Sigma = \left( \begin{smallmatrix}  k_\phi(t_1, t_1') & k_\phi(t_1, t_2')\\ k_\phi(t_2, t_1') & k_\phi(t_2, t_2')\end{smallmatrix} \right)
+ \left( \begin{smallmatrix}  k_\theta(t_1, t_1') & 0\\ 0 & k_\theta(t_2, t_2')\end{smallmatrix} \right),
\end{equation}
is the joint covariance matrix of the two routes around the straight line path - the first matrix indicates the common variance around the straight line deriving from a shared habitual route, and the second matrix indicates the within-path variation, controlling scale of deviation from the other route and smoothness of the trajectory. The conditional probability distribution of one path conditioned on observing the other can be determined by the usual definition of conditional probability:
\begin{equation}
P(x_2(t_2) \mid x_1(t_1)) = \frac{P(x_2(t_2), x_1(t_1))}{P(x_1(t_1)}
\end{equation}
A key finding of our previous study \cite{mann2011oil} was that once a pigeon had completed approximately 5-10 training flights, its future flight paths could be predicted more accurately by incorporating previous training flights into the model. In one sense this simply demonstrated the established phenomenon of habitual route learning through an alternative means. However, in another it provided a crucial step in using flight paths directly to quantitatively test biological hypotheses, since this was the first demonstration that the flight paths itself, rather than an extracted statistic, could be more accurately predicted under one hypothesis than another.

\subsection*{Role of hyperparameters}
The Gaussian process model for pigeon flight path relies on a set of hyperparameters that determine properties of various functions; smoothness of the flight paths and the habitual routes, scales of spatial variation of flight paths around the habitual route and scale of spatial variation of the habitual route around the straight line path. Our approach to these parameters is described in detail in \cite{mann2010poh} and \cite{mann2011oil}, and uses Monte Carlo sampling methods to average over possible values. For the purposes of this paper we assume in the mathematics that these parameters are known, with the caveat that in practice they will be treated using this numerical methods. 

\subsection*{Extension to previous model: Dynamic Time Warping}
In \cite{mann2010poh} we noted that the time-series nature of a recorded flight path posed problems for quantitative prediction of the spatial shape of future paths. A bird could recapitulate a previously memorised route almost exactly, but at different speeds. This new trajectory would appear almost identical to its previous flights when plotted on a map, but very different when the $X$ and $Y$ coordinates were plotted as functions of time. This poses problems when it is solely the spatial structure of the route that interests us. To some degree this problem can be ameliorated by defining a new time coordinate that varies continuously between zero at the start of the flight and one at its completion, as was done is \cite{mann2011oil}. This can compensate for a flight that is uniformly slower or faster than before, but not for one where speed varies significantly along the route (if the bird circles in flight to regain its bearing for example). This is more frequently a problem comparing flight paths from different individuals than flights by the same individual, since speed profiles vary between birds. This problem can be addressed using techniques designed for aligning time series, prominent in voice and video processing, known as Dynamic Time Warping (DTW). This produces an alignment between two times series by adapting the time coordinate at different rates along each series. DTW has been used to compare and cluster trajectories in other fields of movement such as commuting patterns \cite{buchin2011detecting} and forms the basis of the Fr\'{e}chet distance between curves. In this paper we use DTW to align routes by minimising the Fr\'{e}chet distance, prior to calculating their joint and independent probabilities. After alignment we normalise all paths to contain 100 pairs of (x,y) coordinates, using a cubic spline interpolation to resample. Because DTW is impractical for simultaneously aligning for than two sequences \cite{zhou2012generalized} we have restricted ourselves to models that only require alignment of pairs of flights; this is a limitation we will be seeking to address in future work.

\section*{Predicting an untested pigeon's flight path}
Our previous work \cite{mann2011oil} focused on predicting a pigeon's flight paths using its own previously recorded flights from the same release site as training data for a predictive model. Our ultimate goal is to predict a previously untested pigeon's flight path from a novel release site. The next stage towards this goal is to predict the flight path of a trained but previously unrecorded pigeon (that is, not using any data specific to the focal pigeon itself) from a release site that has been used with other birds.

If visual features (or other salient and consistently available navigational cues) are non-uniformly distributed across the landscape then we should expect that multiple birds will tend to navigate along a limited set of flight `corridors' -- a prediction which is qualitatively born out by visual inspection. In this case we should be able to predict the stable, trained flight path of a new bird using data from other individuals.

In \cite{mann2010poh} we developed a model for analysing the flights of birds released in pairs, using a Gaussian process mixture model to predict their flight paths when flying together based on their individual training flights. We can use the same model to make predictions about a new bird from a previously used release site, by predicting that the new bird's flight path will be generated from one or more of the routes we have observed other birds taking. Letting $x_i(t)$ be the last observed route of previously observed bird $i$, and $x^*(t)$ be the as-of-yet unobserved flight path of the new bird, we specify the following mixture model:
\begin{equation}
P(x^*(t) \mid x_1(t), x_2(t), \ldots x_n(t)) = \frac{1}{n} \sum_{i=1}^n P(x^*(t) \mid x_i(t)) 
\end{equation}
where $P(x^*(t) \mid x_i(t))$ is to be evaluated using the Gaussian process model previously developed.

\section*{Methods and data}
We used data from a total 31 trained birds, released from four distinct release sites in the Oxford area (Horspath, n=8; Weston Wood, n=8; Bladon Heath n=7; Church Hanborough, n=8). This data set is identical to that used in \cite{mann2011oil} and the reader is directed to the relevant sections in that study for discussion on data collection protocols. 

For each bird we attempted to predict its final flight of twenty consecutive releases, using a variety of possible predictors to train our Gaussian process model:
\begin{enumerate}[(a)]
\item Only the start and end points.
\item The individual bird's own previous training flight.
\item The last five training flights by this individual (as a mixture model).
\item The final training flights of all \emph{other} birds from this release site (as a mixture model) 
\item The actual final flight path as if it was a previous flight.
\end{enumerate}
where the last item is included to indicate the best we should expect to be able to do in terms of predictive power using this model.

\section*{Results}
\begin{figure}[!h]
\centering
\includegraphics[width= 14cm]{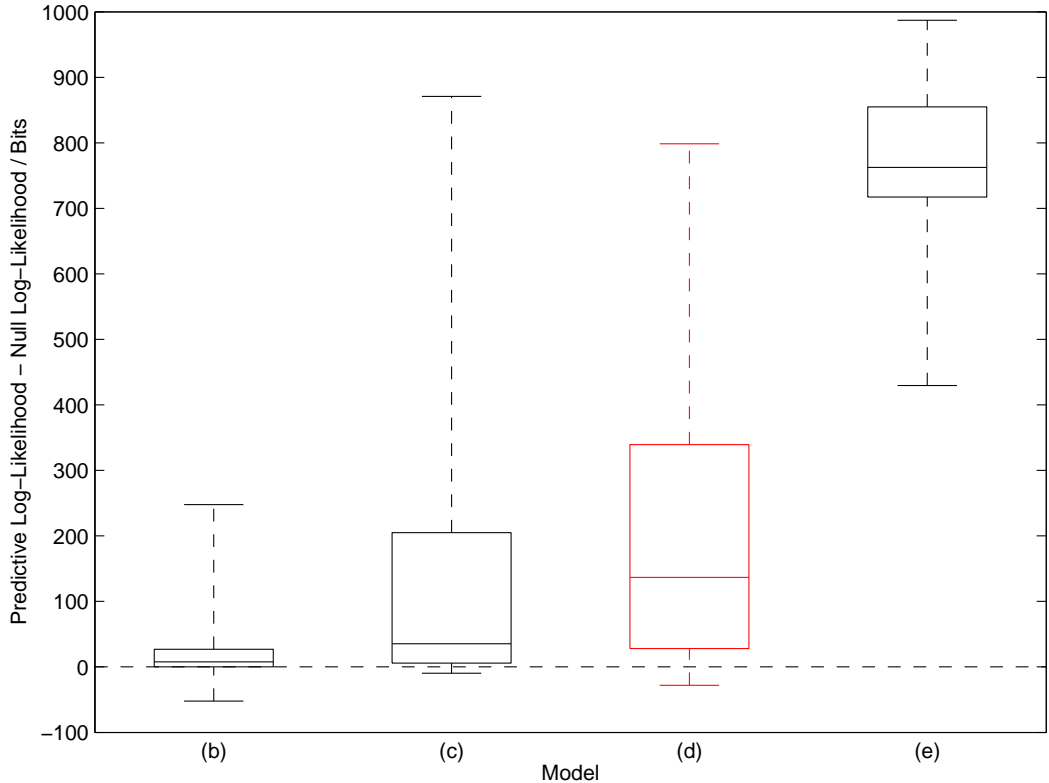}
\caption{The quality of predictions of flight paths, based on different models. The box plot shows the median, interquartile range and maximal range of the log-likelihood of different birds' flight paths for each model, after subtraction of the log-likelihood for the null model (in base 2 logarithms). Highlighted in red is the prediction of a bird's flight route based on the routes of other birds. The results show that other birds' flight paths can successfully be sued to predict a new bird's route, often even more accurately than using the focal bird's own previous flights. These results are aggregated over four distinct release sites; site by site comparisons are shown in Figure \ref{fig:results_sites}}
\label{fig:results}
\end{figure}

\begin{figure}[!h]
\centering
\includegraphics[width= 14cm]{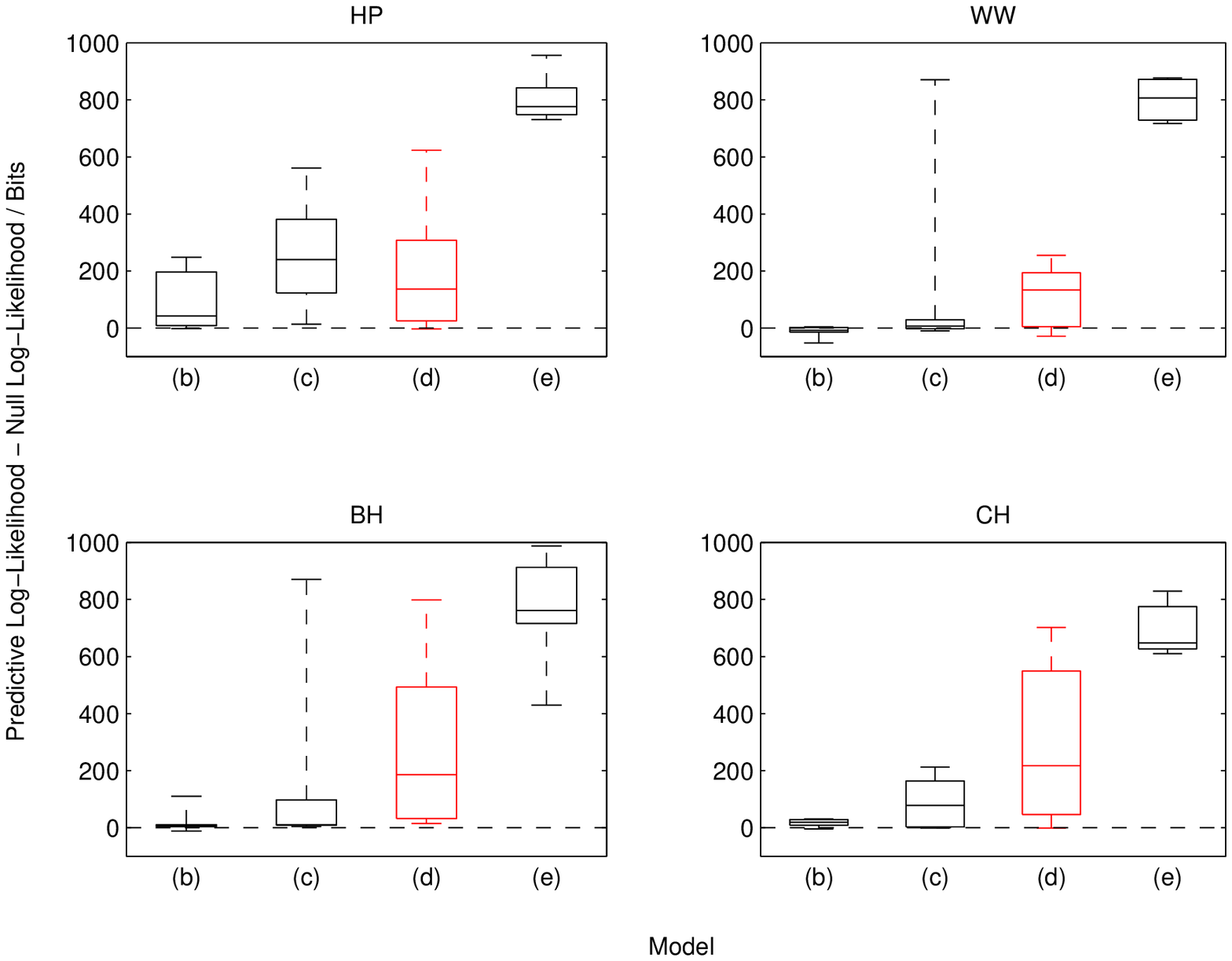}
\caption{Site by site comparison of model predictive accuracy, for four sites: Horspath (HP); Weston Wood (WW); Bladon Heath (BH); and Church Hanborough (CH). Aggregated results in Figure \ref{fig:results} are broadly replicated at each specific site.}
\label{fig:results_sites}
\end{figure}

We evaluated the predictive power of each model listed above in terms of the log-likelihood for the predicted path. In Figure \ref{fig:results} we show these results aggregated over all birds and all sites, with each model log-likelihood shown relative to the simplest hypothesis, i.e. subtracting the log-likelihood under model (a). Highlighted in red is the prediction made using other birds from the same site. The results show that the best predictor, as expected, is the same route as that being predicted. Within the other models the prediction using other birds from the same site (model (d)) actually performs somewhat better than the focal individuals own previous routes, either the single previous flight (model (b)) or a mixture of its previous five routes (model (c)). As shown in Figure \ref{fig:results_sites}, this pattern is broadly replicated across the four different release sites. The key finding is that for all release sites, and in the aggregate, prediction using the previous flights of \emph{other birds} improves on the naive prediction using only the known release and home locations. 

\section*{Discussion}
In this paper we have argued that a predictive model of actual movement trajectories is the best way to judge the accuracy of biological hypotheses regarding the nature of homing navigation. We described the development of such a model from previous work and extended on this to predict the homeward route of pigeons without knowledge of their individual previous training flights. To do this we utilised the theory that salient navigational features, being non-uniformly distributed in space, will create preferred navigational corridors shared by multiple different birds. As such we were able to make predictions of one bird's flight from observations of others preferred routes.

Our results indicated that the flight paths of different individuals could be used to predict the favoured route of an unseen bird, demonstrating that flight paths are predictable not just from previous observing an individual itself, but also across individuals. This is evidence from a predictive model that birds form favoured habitual routes from a specific release site which are not uniformly distributed spatially, but are instead clustered into specific corridors. This is most probably the result of spatially fixed salient navigational cues which are non-uniformly distributed in space, and which are shared in use by multiple individuals. We interpret these as likely to be visual cues, but further development of the model is needed to make this claim more concretely. 

An interesting feature of our results is that using other birds observed flight paths to predict a new individual's route often resulted in improved predictions over using the focal individuals own previous flight records. This could be because even experienced birds will sometimes change their flight routes over time, or even switch between two alternative routes. In these cases the use of training data from other birds may capture this wider variety of possible route choices better than the most recent flights by the same bird. 

Going forward, the next ultimate goal of this progression is to be able to predict these flight corridors from a new site before we observe \emph{any} bird released there. In the familiar area, where pigeons are expected to use primarily visual information for navigation, this means constructing an explicit link between the visual landscape, presumably using aerial images as a proxy, and the flight generating model. Potential ways to do this include using experienced researchers to view aerial images and suggest routes, which could be used as training data for a model as in this paper, or algorithmically generating putative landmarks from the aerial image, perhaps using local image complexity or similar measures as in \cite{mann2014landscape}, and subsequently generating simulated paths that link these potential waypoints. 

\subsection*{Acknowledgments}
I would like to thank the participants of the Dagstuhl seminar `Geometric and Graph-based Approaches to Collective Motion' whose thoughtful contributions during discussion groups led me to the literature on Dynamic Time Warping and reinvigorated my interest in this area.
\FloatBarrier
\bibliographystyle{ieeetr}
\bibliography{../bibfiles/pigeon}

\begin{thebibliography}{10}

\bibitem{braithwaite1991vfl}
V.~A. Braithwaite and T.~Guilford, ``Viewing familiar landscapes affects pigeon
  homing,'' {\em Proc. Roy. Soc. B}, vol.~245, pp.~183--186, 1991.

\bibitem{mann2011oil}
R.~Mann, R.~Freeman, M.~Osborne, R.~Garnett, C.~Armstrong, J.~Meade, D.~Biro,
  T.~Guilford, and S.~Roberts, ``Objectively identifying landmark use and
  predicting flight trajectories of the homing pigeon using gaussian
  processes,'' {\em Journal of The Royal Society Interface}, vol.~8, no.~55,
  pp.~210--219, 2011.

\bibitem{biro2004frl}
D.~Biro, J.~Meade, and T.~Guilford, ``Familiar route loyalty implies visual
  pilotage in the homing pigeon,'' {\em Proc. Natl. Acad. Sci. U.S.A.},
  vol.~101, no.~50, pp.~17440--17443, 2004.

\bibitem{meade2005hpd}
J.~Meade, D.~Biro, and T.~Guilford, ``Homing pigeons develop local route
  stereotypy,'' {\em Proc. Roy. Soc. B}, vol.~272, pp.~17--23, 2005.

\bibitem{wiltschko2007hfp}
R.~Wiltschko, I.~Schiffner, and B.~Siegmund, ``Homing flights of pigeons over
  familiar terrain,'' {\em Anim. Behav.}, vol.~74, no.~5, pp.~1229--1240, 2007.

\bibitem{schiffner2014pigeon}
I.~Schiffner and R.~Wiltschko, ``Pigeon navigation: different routes lead to
  frankfurt,'' {\em PLoS ONE}, 2014.

\bibitem{lipp2004pha}
H.~P. Lipp, A.~L. Vyssotski, D.~P. Wolfer, S.~Renaudineau, M.~Savini,
  G.~Tr\"{o}ster, and G.~Dell'Omo, ``Pigeon homing along highways and exits,''
  {\em Curr. Biol.}, vol.~14, pp.~1239--1249, 2004.

\bibitem{lau2006aed}
K.~Lau, S.~Roberts, D.~Biro, R.~Freeman, J.~Meade, and T.~Guilford, ``An
  edge-detection approach to investigating pigeon navigation,'' {\em J. Theor.
  Biol.}, vol.~239, no.~1, pp.~71--78, 2006.

\bibitem{mann2014landscape}
R.~P. Mann, C.~Armstrong, J.~Meade, R.~Freeman, D.~Biro, and T.~Guilford,
  ``Landscape complexity influences route-memory formation in navigating
  pigeons,'' {\em Biology letters}, vol.~10, no.~1, p.~20130885, 2014.

\bibitem{mann2013msi}
R.~P. Mann, A.~Perna, R.~Garnett, J.~E. Herbert-Read, D.~J.~T. Sumpter, and
  A.~J.~W. Ward, ``Multi-scale inference of interaction rules in animal groups
  using bayesian model selection,'' {\em PLOS Computational Biology}, vol.~9,
  no.~3, p.~e1002961, 2013.

\bibitem{mann2013tdo}
R.~P. Mann, J.~Faria, D.~J.~T. Sumpter, and J.~Krause, ``The dynamics of
  audience applause,'' {\em J R Soc Interface}, p.~20130466, 2013.

\bibitem{mann2013humbugs}
R.~P. Mann, J.~E. Herbert-Read, Q.~Ma, L.~A. Jordan, D.~J. Sumpter, and
  A.~J.~W. Ward, ``A model comparison reveals dynamic information drives the
  movements of humbug damselfish (dascyllus aruanus).,'' {\em Journal of the
  Royal Society Interface}, 2013.

\bibitem{reece2011gps}
S.~Reece, R.~Mann, I.~Rezek, and S.~Roberts, ``Gaussian process segmentation of
  co-moving animals,'' in {\em AIP Conference Proceedings-American Institute of
  Physics}, vol.~1305, p.~430, 2011.

\bibitem{rasmussen2006gpf}
C.~E. Rasmussen and C.~K.~I. Williams, {\em Gaussian Processes for Machine
  Learning}.
\newblock The M.I.T Press, 2006.

\bibitem{mann2010poh}
R.~P. Mann, {\em Prediction of Homing Pigeon Flight Paths using Gaussian
  Processes}.
\newblock PhD thesis, University of Oxford, 2010.

\bibitem{buchin2011detecting}
K.~Buchin, M.~Buchin, J.~Gudmundsson, M.~L{\"o}ffler, and J.~Luo, ``Detecting
  commuting patterns by clustering subtrajectories,'' {\em International
  Journal of Computational Geometry \& Applications}, vol.~21, no.~03,
  pp.~253--282, 2011.

\bibitem{zhou2012generalized}
F.~Zhou and F.~De~la Torre, ``Generalized time warping for multi-modal
  alignment of human motion,'' in {\em Computer Vision and Pattern Recognition
  (CVPR), 2012 IEEE Conference on}, pp.~1282--1289, IEEE, 2012.

\end{thebibliography}

\end{document}